*Article*

# Switch Elements with S-Shaped Current-Voltage Characteristic in Models of Neural Oscillators

**Petr Boriskov * and Andrei Velichko**

Institute of Physics and Technology, Petrozavodsk State University, 31 Lenina str., Petrozavodsk 185910, Russia

* Correspondence: boriskov@petrsu.ru; Tel.: +79062067901

**Abstract:** In this paper, we present circuit solutions based on a switch element with the S-type I–V characteristic implemented using the classic FitzHugh–Nagumo and FitzHugh–Rinzel models. Using the proposed simplified electrical circuits allows the modeling of the integrate-and-fire neuron and burst oscillation modes with the emulation of the mammalian cold receptor patterns. The circuits were studied using the experimental I–V characteristic of an $NbO_2$ switch with a stable section of negative differential resistance (NDR) and a $VO_2$ switch with an unstable NDR, considering the temperature dependences of the threshold characteristics. The results are relevant for modern neuroelectronics and have practical significance for the introduction of the neurodynamic models in circuit design and the brain–machine interface. The proposed systems of differential equations with the piecewise linear approximation of the S-type I–V characteristic may be of scientific interest for further analytical and numerical research and development of neural networks with artificial intelligence.

## 1. Introduction

Mathematical modeling of the nerve cell behavior, based on a detailed understanding of the functioning mechanisms of a neuron as a biological object, allows a quantitative prediction of neuron behavior. The parameters of such models reflect a clear biophysical meaning and can be measured, or estimated, in real bio-experiments. However, such modeling does not often allow the possibility of analytical research. To study the behavior of neurodynamic systems, simplified models are used, in particular, the FitzHugh–Nagumo [1–7] and FitzHugh–Rinzel [8] models. Such circuits, on the one hand, should preserve the general principle of generating neural signals, and, on the other hand, should have simpler circuit solutions in terms of implementing cognitive technologies, for example, spike (SNN) [9] and oscillatory (ONN) neural networks [10–16]. Spike-based models of neurons based on silicon technology have recently become popular in the academic literature, and a good overview of these models is given in [17]. The most prominent models include a "Tau-Cell neuron" model, based on the first-order low-pass filter [18], an Axon–Hillock circuit [19], and an Izhikevich neuron circuit [20], which implement a variety of spiking behaviors, such as regular spiking, spike-frequency adaptation, and bursting. Therefore, the development of neurodynamic models in circuit design is an important task of modern neuroelectronics, in particular, in the field of brain–machine interface.

An electrical switching (ES) is an abrupt, significant, and reversible change in the conductivity of an electric element under the applied electric field or a flowing current. Current-voltage characteristic (I–V) of the element contains areas with negative differential resistance (NDR), created by the positive feedback of current (S-type I–V) or voltage (N-type I–V) [21–24]. However, this feedback is internal and is not created by external circuit elements, like in radio circuits with operational amplifiers.

A trigger diode is a typical element with an S-type I–V characteristic and the effect of electrical switching. It is formed by three consecutive p-n junctions, where the current instability is caused by the injection of minority carriers into the middle p-n junction [25]. Modern silicon technology allows the creation of dynistor elements with the S-type I–V characteristic of a wide range of voltages and currents, characterized by stability and low noise level.

Among non-silicon materials with ES effect, amorphous semiconductors and transition metal oxides material classes can be highlighted. In the amorphous semiconductors material class, the phenomena of electrical instability are mainly studied in chalcogenide glasses (CGSs), which contain elements of VI group (Se, S, and Te) and elements of the type Si, Ge, Bi, Sb, As, and P). In CGSs, the switching can have the stable part of NDR (Figure 1a), the unstable part of NDR (Figure 1b), and the memory effect (Figure 1c) [21]. Among transition metal oxides, S-type threshold switching is observed in metal-oxide-metal (MOM) structures based on $Nb_2O_5$, $NbO_2$, $TiO_2$, $VO_2$, $Ta_2O_5$, Fe oxide, and some other oxides [22,24]. In at least two oxides ($NbO_2$ and $VO_2$), ES is caused by the metal-insulator phase transition (MIT). Recently, the switching effect has been studied in high-temperature superconductivity (HTSC) [26], colossal magnetoresistance (CMR) manganites, and the heterostructures based on them [27], as well as in various carbon-containing materials, including fullerenes and nanotubes [28,29].

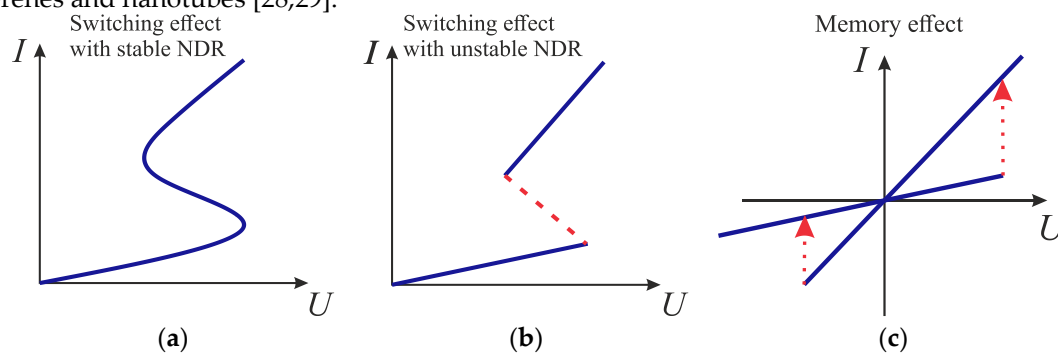

**Figure 1.** S-type I–V characteristic with stable NDR segments (**a**), unstable NDR segments, (**b**) and with bipolar memory effect (**c**).

In this paper, we propose a set of circuit solutions based on a generalized switch element with S-type I–V characteristic that implements the classic FitzHugh–Nagumo and FitzHugh–Rinzel models. We study circuit operations using the experimental I–V characteristic with a stable ($NbO_2$ switch) and an unstable ($VO_2$ switch) NDR sections. In addition, we propose simplified electrical circuits, allowing the ability to simulate an integrate-and-fire neuron and burst oscillation mode with the emulation of mammalian cold receptor patterns.

## 2. Materials and Methods

### 2.1. S-Type Switch Models Controlled by Current and Voltage

Using an example of a planar $VO_2$ switch, a typical S-type I–V characteristic with an unstable NDR [30] is presented in Figure 2a. Electrical instability causes the presence of a high-resistance (OFF) and low-resistance (ON) branches with threshold voltages (currents) of switching on ($U_{th}$, $I_{th}$) and switching off ($U_h$, $I_h$).

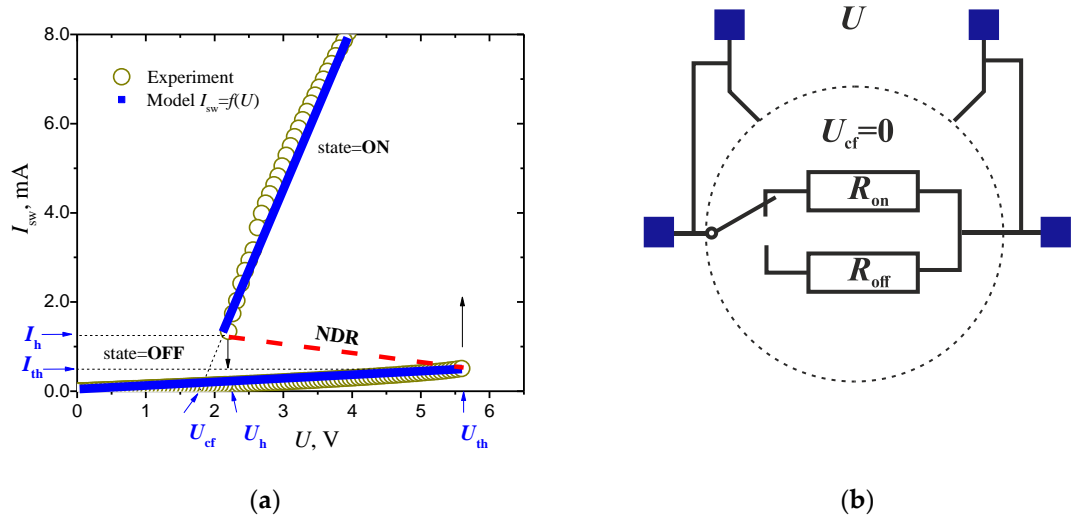

**Figure 2.** Experimental [30] and model I–V characteristic of a planar VO$_2$ switch with an unstable NDR (**a**) and an electrical key diagram for a model I–V characteristic with $U_{cf}=0$ (**b**). The VO$_2$ switch parameters for $I_{sw}(U)$ (1-2) are: $U_{th}$ = 5.64 V, $U_h$ = 2.12 V, $U_{cf}$ = 1.754 V, $R_{off}$ = 10742 Ω and $R_{on}$ = 276 Ω.

If both branches are approximated by straight lines with differential resistances $R_{off}$ and $R_{on}$, the model I–V characteristic has the form

$$I_{sw}(U) \approx \begin{cases} U/R_{on}, & \text{state = OFF} \\ (U - U_{cf})/R_{off}, & \text{state = ON} \end{cases} \quad (1)$$

where $U_{cf}$ — (cutoff voltage) residual voltage of low-resistance (ON) section. Switching in Equation (1) between OFF and ON states is implemented as follows:

$$\text{state} = \begin{cases} \text{OFF}, & \text{if (state = ON) and } (U < U_h) \\ \text{ON}, & \text{if (state = OFF) and } (U > U_{th}) \end{cases} \quad (2)$$

Equations (1) and (2) characterize the switch element and which state is determined by the voltage on the element (Figure 2b). We have already used a similar equation for modeling and analyzing experimental data on the dynamics of VO$_2$ oscillators [16].

Using the example of NbO$_2$ sandwich switch [31], the S-type I–V characteristic with stable NDR is presented in Figure 3.

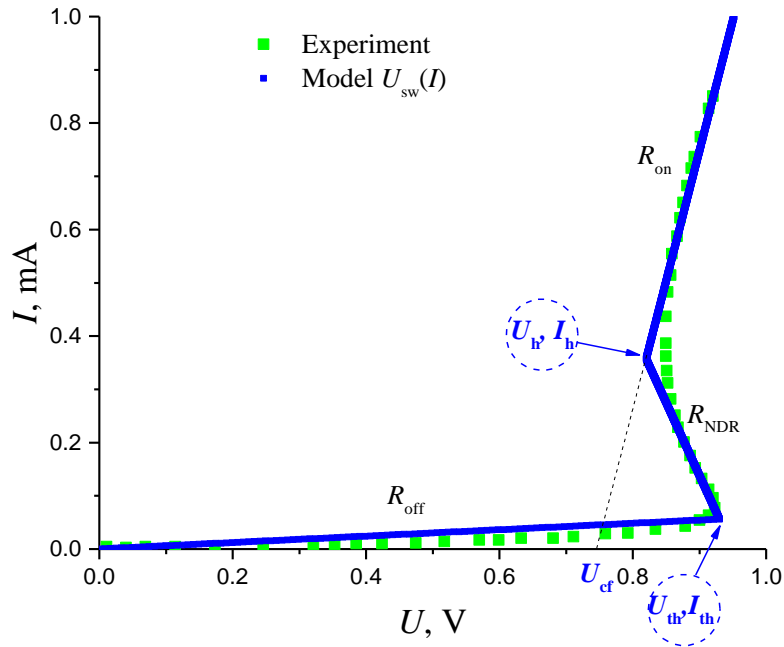

**Figure 3.** Experimental [31] and the model I–V characteristic of a NbO$_2$ sandwich switch with a stable NDR. Parameters of the NbO$_2$ switch for $U_{sw}(I)$ (3-4): $I_{th}$ = 56 μA, $I_h$ = 357 μA, $U_{th}$ = 0.93 V, $U_h$ = 0.82 V, $U_{cf}$ = 0.747 V, $R_{NDR}$ = –365 Ω, $R_{off}$ = 16.61 kΩ, and $R_{on}$ = 204.5 Ω.

The I–V characteristic with a stable NDR is an unambiguous and continuous voltage-to-current function that characterizes a current-controlled switch element. Its modeled piecewise linear approximation by straight lines can be written as:

$$U_{sw}(I) = \frac{1}{2} \cdot \left[ (R_{on} + R_{off}) \cdot I + (R_{NDR} - R_{off}) \cdot (|I - I_{th}| - I_{th}) - (R_{NDR} - R_{on}) \cdot (|I - I_h| - I_h) \right] \quad (3)$$

with negative resistance in the NDR section

$$R_{NDR} = \frac{U_h - U_{th}}{I_h - I_{th}} \quad (4)$$

In this way, we have identified two types of elements with a stable and unstable NDR on the I–V characteristic, described by a piecewise linear approximation determined by five independent parameters. Equation (3), describing the S-type I–V characteristic with stable NDR, has been formulated by us for the first time, although a linear approximation, using the module function, has been used earlier, for example, to describe the I–V characteristic of Chua's diode [32].

Not every circuit can function with elements of both types. For example, when inductors are connected in series with a voltage-controlled element (the I–V characteristic with an unstable NDR, see Figure 2a,b), the solution may not exist. During the numerical simulation of the circuit, the solution became more unstable with a decrease in the calculation step, since inductance operation does not imply an instantaneous change in current. Therefore, in the diagrams below, we indicate the type of switch with the designation in the form of color resistors (Figure 4), controlled by current (Figure 4a) or voltage (Figure 4b). If the circuit can work with both types of switches, the combined designation is used (see Figure 4c).

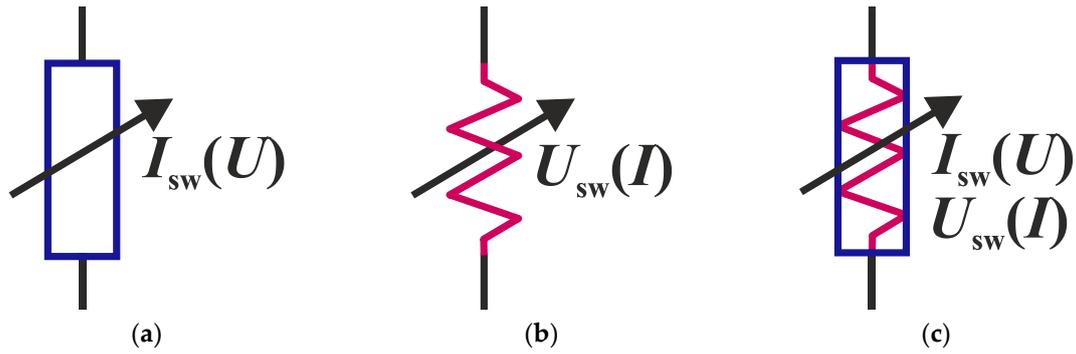

**Figure 4.** Designation for a switching element controlled by a voltage (**a**), current (**b**) and, optionally, current or voltage (**c**).

The circuits presented in the article were implemented using the MathCAD and LTSpice software packages. With the same parameters of the circuits, the simulation results in both software packages were identical. LTSpice source files and NDR modeling code are available in Supplementary Materials.

*2.2. Relaxation Oscillator*

An element with an S-shaped NDR can be included in an active RC circuit (Figure 5). If the supply current $I_0$ of the circuit is in the NDR range

$$I_{th} < I_0 < I_h \qquad (5)$$

the circuit generates self-oscillations of the relaxation type, similar to the oscillations of the classical auto-oscillator on a Pearson–Anson neon lump [33].

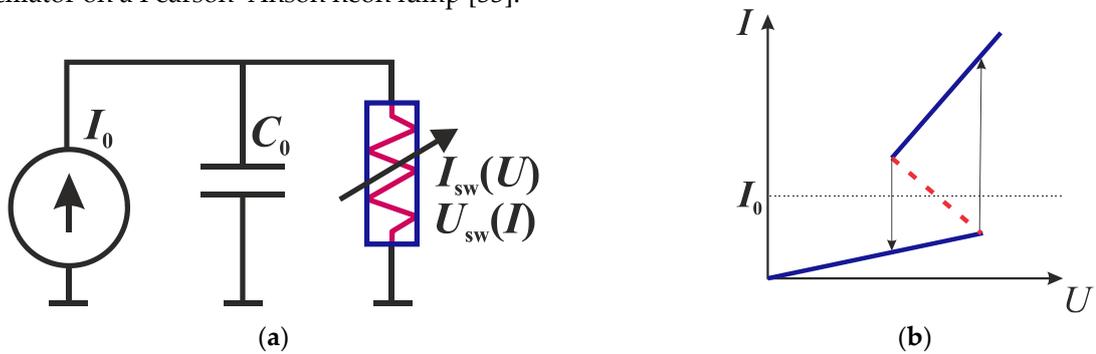

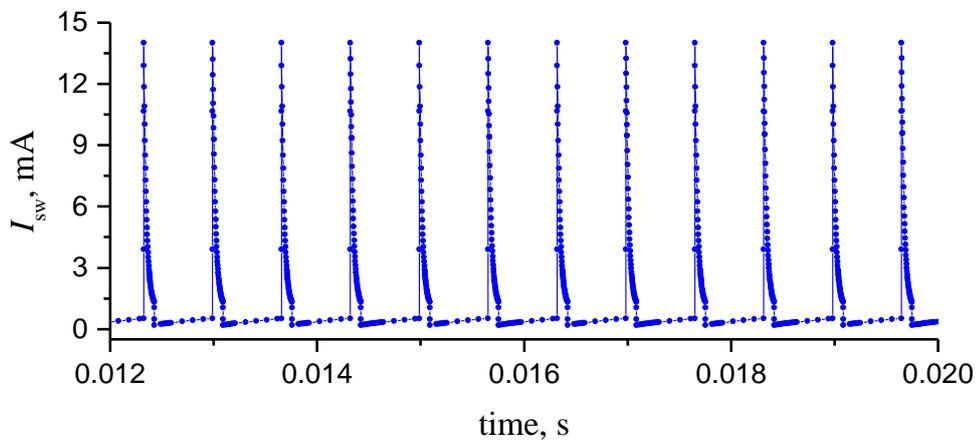

(c)

**Figure 5.** Relaxation oscillator circuit (**a**) and the load characteristic with the operating point in the NDR region (**b**). Current oscillations at $I_0$ = 1 mA, $C_0$ = 100 nF, and VO$_2$ switch (for the I–V characteristic with unstable NDR, see Figure 2a) (**c**).

By modifying the basic circuit, it is possible to compile various neural models into one. As there are no inductive elements in the illustrated circuit, there is no difference what type of S-element NDR is used to generate relaxation oscillations.

*2.3. FitzHugh–Nagumo and FitzHugh–Rinzel Models*

The Hodgkin–Huxley model (HHM) served as the theoretical foundation for the ionic mechanisms involved in the excitation and inhibition in the peripheral and central parts of the nerve cell membrane [20,34–42]. The FitzHugh–Nagumo (FN) model [38] is the first simplified version of HHM, which only allows for the modeling of self-excitation and restoration of the membrane potential through positive and negative feedback.

The following equations describe the model in a dimensionless form:

$$\frac{dG}{d\tau} = G - \frac{G^3}{3} - W + I_{ext},$$
$$\tau_0 \frac{dW}{d\tau} = G + \alpha - \beta W \quad (6)$$

where $G(\tau)$ is a variable describing the dynamics of membrane potential with a current $I_{ext}(\tau)$, $W(\tau)$ is a recovery variable, $\alpha$, $\beta$, and $\tau_0$ are experimentally determined parameters. In this model, with a cubic nonlinearity ($F_q(G) = G - G^3/3$), a bi-stability regime can be observed, when the gravity regions of two stable equilibrium states are separated by a saddle [41,43].

The FitzHugh–Rinzel (FR) model is the development of the FitzHugh–Nagumo modified generator [44], where the additional slowly changing variable $Q(\tau)$ is added. The system of equations described in the FR model in the dimensionless form is as follows [8]:

$$\frac{dG}{d\tau} = G - \frac{G^3}{3} - W + Q + I_{ext},$$
$$\tau_0 \frac{dW}{d\tau} = G - \beta W, \quad \tau_1 \frac{dQ}{d\tau} = \gamma - G - \chi Q \quad (7)$$

The resulting 3D model, with five parameters ($\tau_0$, $\tau_1$, $\beta$, $\chi$, and $\gamma$), helps to study specific reactions of the nerve cell, such as the regular and chaotic generation of bursts [45]. For the application in ONN and SNN, this model enables the frequency and phase coding of information, not only with the help of a spikes sequence, but also with bursts synchronization.

**3. Results**

*3.1. FitzHugh–Nagumo Model Based on a Current-Controlled Switching S-Element*

we added a reactive element to the circuit of the auto-relaxation oscillator (Figure 5a). The reactive element added was inductance (*L*), and, in parallel to the supply current $I_0$, we added the resistance element ($R_0$), as shown in Figure 6. The inductance introduces an additional degree of freedom to the system (inductance current), which determines the charging–discharging process of capacity ($C_0$) during switching of the NDR element. In addition, the circuit includes the sources of direct current $I_{b0}$ and alternating voltage:

$$U_b(t) = U_{b0} + U_{in}(t) \quad (8)$$

where $U_{in}(t)$ is the input (stimulating) signal relative to the constant bias $U_{b0}$. The use of these sources to generate oscillations in the circuit of Figure 6, as well as in the subsequent circuits (Section 3.2 and 3.3), is optional, but they are necessary for a general interpretation of the circuits of FitzHugh–Nagumo (Equation (6)) and FitzHugh–Rinzel (Equation (7)) models.

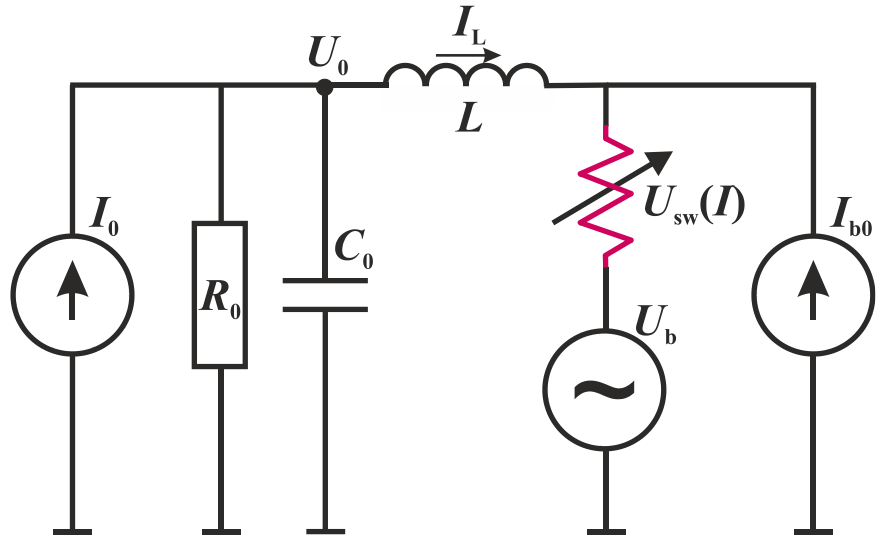

**Figure 6.** Oscillator circuit implementing the FitzHugh–Nagumo model.

The mathematical model of the circuit in Figure 6, expressed in the form of a system of equations based on Kirchhoff's laws, has the following form:

$$C_0 \frac{dU_0}{dt} = I_0 - \frac{U_0(t)}{R_0} - I_L(t),$$

$$L \frac{dI_L}{dt} = U_0(t) - U_b(t) - U_{sw}\left(I_L(t) + I_{b0}\right)$$

(9)

where $I_L$ and $U_0$ are the inductive current and voltage on the capacitor $C_0$, respectively.

The midpoint parameters of the NDR of S-type I–V characteristic are:

$$I_{mp} = \frac{I_h + I_{th}}{2}, \ U_{mp} = \frac{U_h + U_{th}}{2}, \ R_{mp} = \frac{U_{mp}}{I_{mp}}$$

(10)

After the transition to dimensionless variables,

$$\tau \equiv t \frac{R_{mp}}{L}, \ G \equiv -\frac{I_L}{I_{mp}}, \ W \equiv \frac{U_0}{U_{mp}}, \ I_{ext} \equiv \frac{U_{in}}{U_{mp}},$$

(11)

the system in Equation (9) is transformed into the FitzHugh–Nagumo model from Equation (6) with the parameters: $\tau_0 = \frac{R_{mp}^2 C_0}{L}$, $\alpha = \frac{I_0}{I_{mp}}$, $\beta = \frac{R_{mp}}{R_0}$, and a cubic nonlinearity is replaced with a piecewise linear function:

$$F_{pw}(G) = \frac{U_{b0}}{U_{mp}} + \frac{U_{sw}\left(-G \cdot I_{mp} + I_{b0}\right)}{U_{mp}}.$$

(12)

In Equation (12), $U_{sw}(I)$ is a function of the argument $I = -G \cdot I_{mp} + I_{b0}$, and is determined from the model I–V characteristic in Equation (3).

Figure 7 depicts the functions $F_q(G)$ and $F_{pw}(G)$ at $U_{b0} = -U_{mp}$ and $I_{b0} = I_{mp}$. In this case, the midpoint of the NDR segment for the $F_{pw}(G)$ function is shifted to the origin, and it corresponds to the best approximation of the $F_{pw}(G)$ function using cubic nonlinearity $F_q(G)$.

Based on Equation (11), the inductive current $I_L$ models the behavior of the membrane potential G, and the voltage on the capacitor $U_0$ reflects the slow recovery potential W. The input signal $U_{in}(t)$ is analogous to the input current $I_{ext}(t)$ of the neuron in Equations (6) and (7).

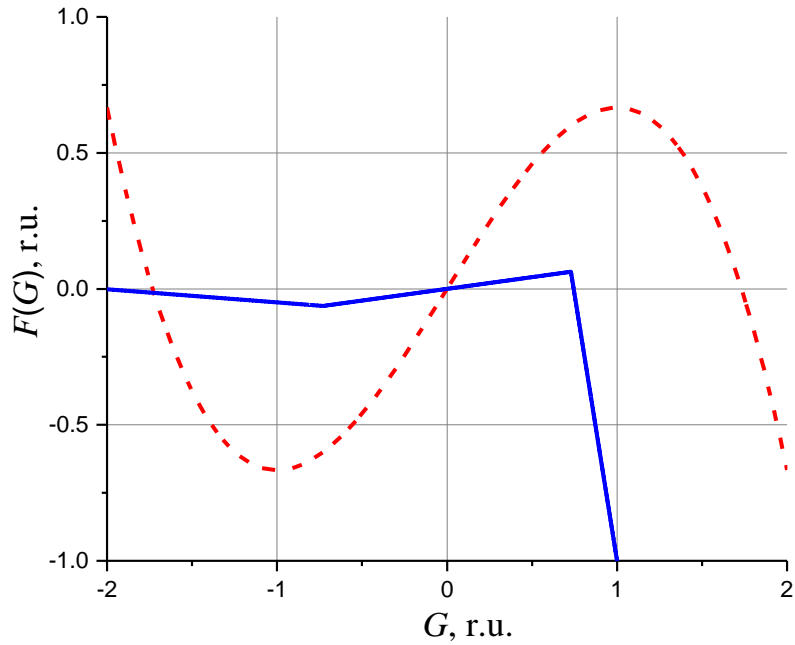

**Figure 7.** The functions of potential $F(G)$ for cubic nonlinearity $F_q(G)$ (dash, red) and piecewise linear approximation of $F_{pw}(G)$ from Equation (12) at $I_{b0} = I_{mp}$ and $U_{b0} = -U_{mp}$ (solid, blue), using the example of I–V characteristic of NbO$_2$ switch (Figure 3).

Figure 8 depicts the regular oscillations of inductance current $I_L(t)$ and voltage $U_0(t)$ at capacitance $C_0$ and voltage $U_{sw}(t)$ at the switch, calculated using the example of the numerical solution of Equation (9) for a switching structure with the experimental I–V characteristic of the S-type, presented in Figure 3. Similar oscillograms can be obtained by modeling the FN circuit in LTspice (see Supplementary Materials).

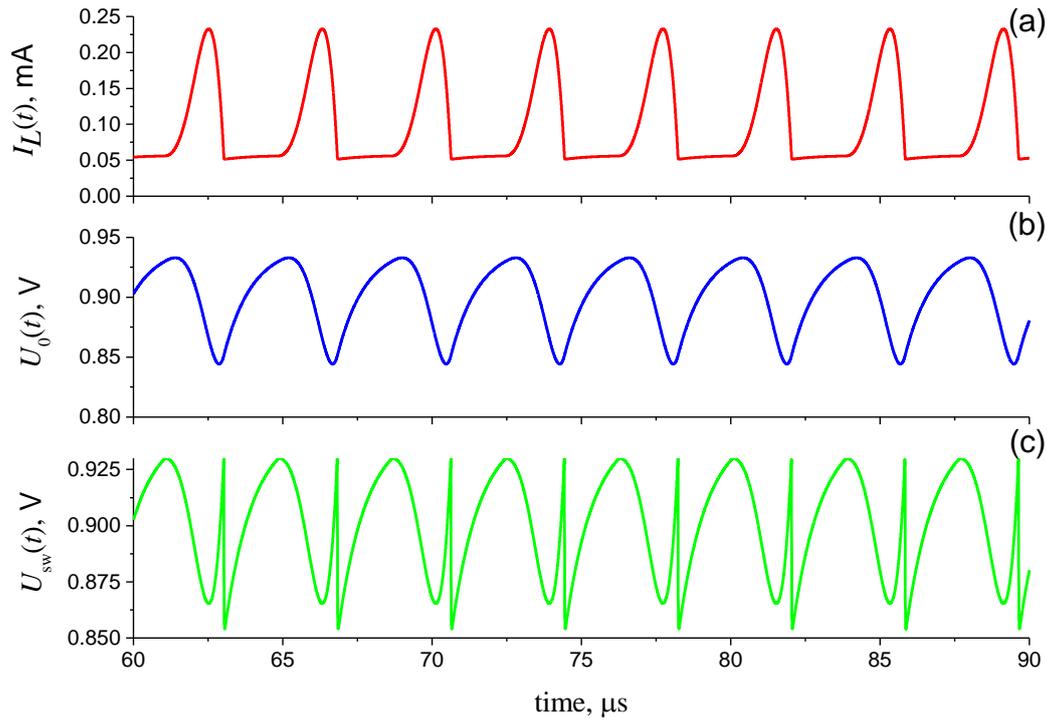

**Figure 8.** The time dependences of the induced current $I_L(t)$ (**a**), the voltage $U_0(t)$ on the capacitance $C_0$ (**b**) and the voltage $U_{sw}(t)$ on the switch (**c**) in the circuit (Figure 6), using the switching S-element (Figure 3). Electrical circuit parameters: $L = 0.1$ mH, $C_0 = 1$ nF, $R_0 = 1$ kΩ, $I_0 = 1$ mA, $I_{b0} = 0$ A, $U_{b0} = 0$ V, and $U_{in}(t) = 0$ V.

*3.2. FitzHugh–Rinzel Model Based on a Current-Controlled Switching S-Element*

Since the FitzHugh–Rinzel model is obtained by adding one more variable to the FitzHugh–Nagumo equations, two options for upgrading the circuit in Figure 6 can be suggested, which is adding an RC filter of either high (Figure 9a) or low frequencies (Figure 9b).

In the first case (Figure 9a), the sequential $R_1C_1$ link below the switch generates a voltage signal $U_1(t)$ on the capacitance $C_1$, which is modulated by the charging–discharging processes. The mathematical model of this circuit (Figure 9a) is represented by a system of equations:

$$\begin{aligned} C_0 \frac{dU_0}{dt} &= I_0 - \frac{U_0(t)}{R_0} - I_L(t), \\ C_1 \frac{dU_1}{dt} &= I_L(t) - \frac{U_1(t)}{R_1} + I_{b0}, \\ L \frac{dI_L}{dt} &= U_0(t) - \left(U_1(t) + U_b(t) + U_{sw}(I_L(t) + I_{b0})\right) \end{aligned} \quad (13)$$

Using the parameters of the S-type I–V characteristic in Equation (11), the system of equations in Equation (13) after the transition to dimensionless variables has the form:

$$\tau \equiv t \frac{R_{mp}}{L}, \; G \equiv -\frac{I_L}{I_{mp}}, \; W \equiv \frac{R_1 I_{b0} - U_1}{U_{mp}}, \\ Q \equiv -\frac{U_0}{U_{mp}}, \; I_{ext} \equiv \frac{R_1 I_{b0} + U_{in}}{U_{mp}} \quad (14)$$

The system is converted to the FitzHugh–Rinzel model from Equation (7) with a piecewise linear function from Equation (12) and with the added parameters:

$$\tau_0 = \frac{R_{mp}^2 C_1}{L}, \quad \tau_1 = \frac{R_{mp}^2 C_0}{L}, \quad \beta = \frac{R_{mp}}{R_1}, \quad \chi = \frac{R_{mp}}{R_0}, \quad \gamma = -\frac{I_0}{I_{mp}} \quad (15)$$

Therefore, fluctuations of the induction current model the temporal dependence of the membrane potential of the neuron $G$, and the oscillations of the $U_0$ and $U_1$ voltages model the recovery potentials $Q$ and $W$, respectively. In Equation (15), all parameters are positive, except for $\gamma < 0$, similar to Rinzel's work [8] with cubic nonlinearity.

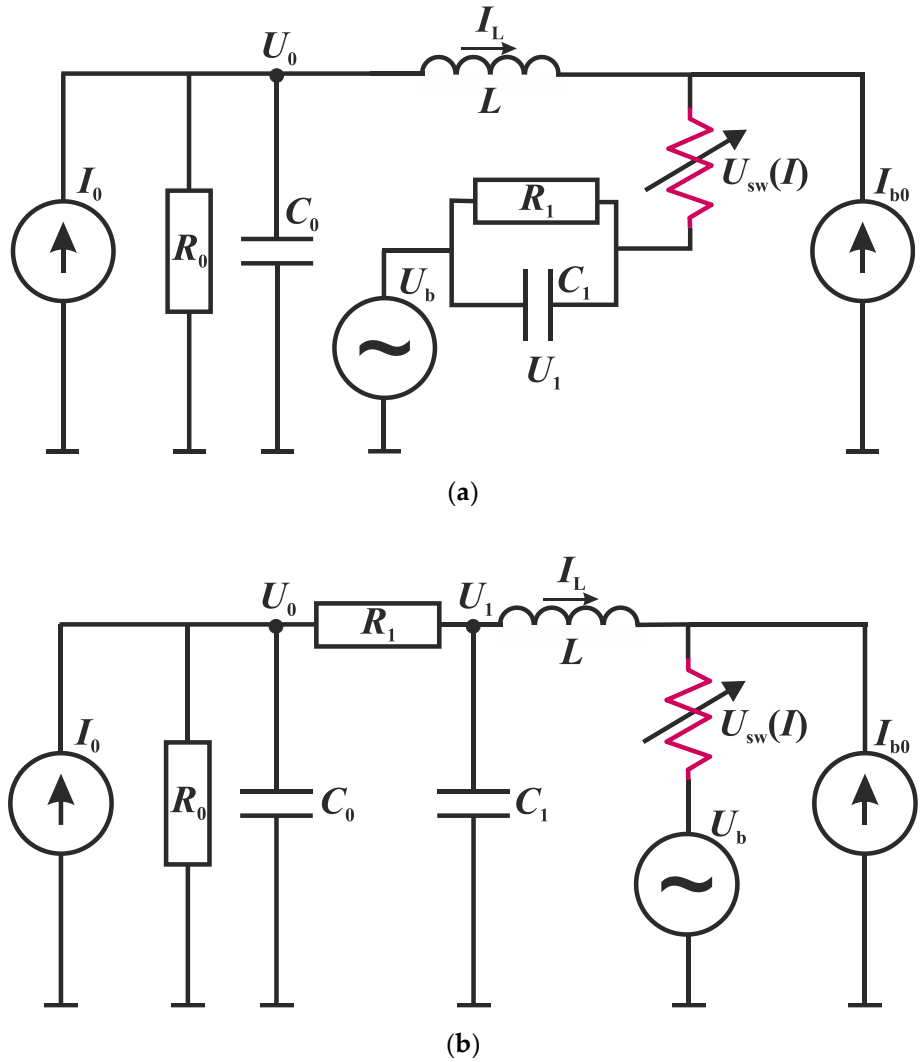

(a)

(b)

**Figure 9.** An oscillator circuit that implements the FitzHugh–Rinzel model with a high-pass $R_1C_1$-filter connected in series to the switch (**a**) and a low-pass $R_1C_1$-filter connected between the capacitor $C_0$ and the inductance $L$ (**b**).

Figure 10 demonstrates the bursts oscillations of the inductance current $I_L(t)$ and voltages $U_0(t)$, $U_1(t)$, and $U_{sw}(t)$ using the example of the numerical solution of Equation (13) for a switching structure with the experimental S-type I–V characteristic presented in Figure 3. In the calculations, similarly to the FitzHugh–Nagumo model in Figure 8, the S-type I–V characteristic's shift to the origin is not used and there is no input (stimulating) signal: $I_{b0} = 0$ A, $U_{b0} = 0$ V, and $U_{in}(t) = 0$ V.

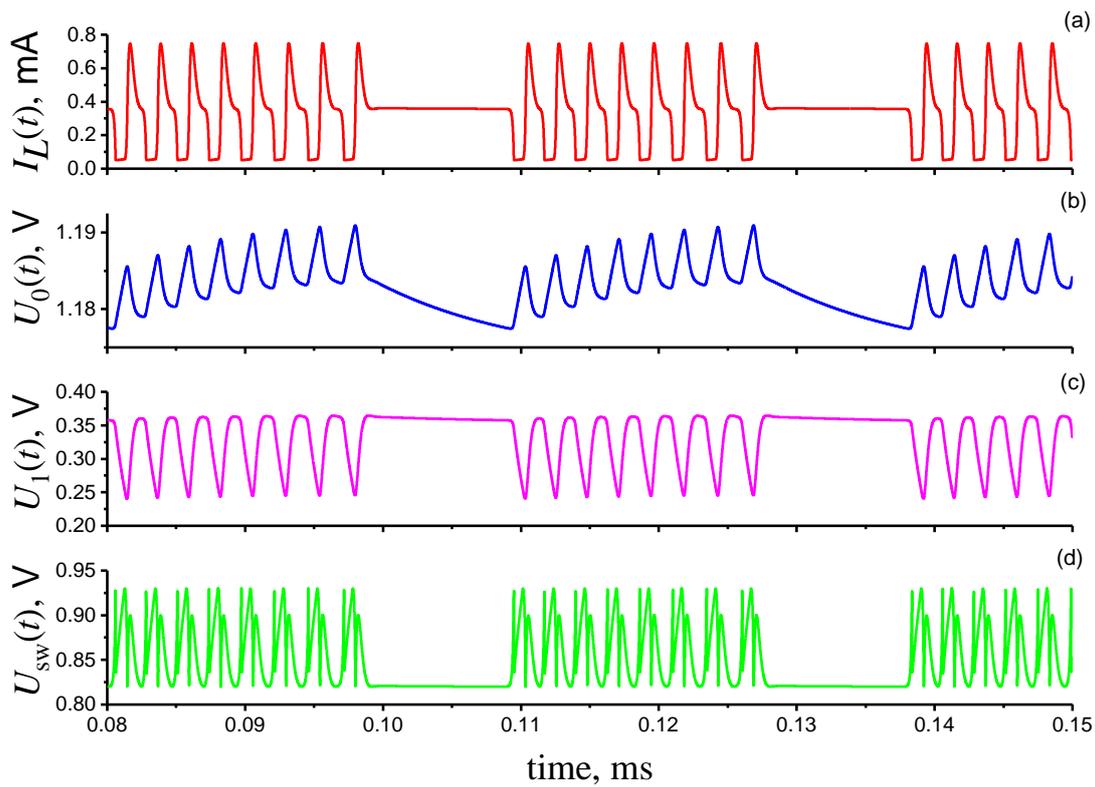

**Figure 10.** The time dependences of the induction current $I_L(t)$ (**a**), voltage $U_0(t)$ at capacitance $C_0$ (**b**), voltage $U_1(t)$ at capacitance $C_1$ (**c**), and voltage $U_{sw}(t)$ at switch (**d**) in the circuit in Figure 9a with a switching S-element (Figure 3). Electrical circuit parameters: $L$ = 0.025 mH, $C_0$ = 30 nF, $C_1$ = 1.75 nF, $R_0$ = 0.5 kΩ, $R_1$ = 1 kΩ, $I_0$ = 2.7 mA, $I_{b0}$ = 0 A, $U_{b0}$ = 0 V and $U_{in}(t)$ = 0 V.

In the second case of the FitzHugh–Rinzel circuit (Figure 9b), the $R_1C_1$ integrator (low-pass filter) is included in the oscillating circuit and acts as a voltage modulator $U_1(t)$ between capacitance $C_0$ and inductance $L$ due to charging–recharging capacitance $C_1$. Appendix A demonstrates how this circuit is converted to the FitzHugh–Rinzel model (Equation (7)) after a linear transformation of variables.

*3.3. Alternative Neural-Like Circuits Based on a Switching S-Element*

Alternative neural-like circuits originate from a relaxation generator circuit (Figure 5a). Additional capacity ($C_1$) is connected in parallel with either the S-switch (Figure 11a) or inductance (Figure 11b).

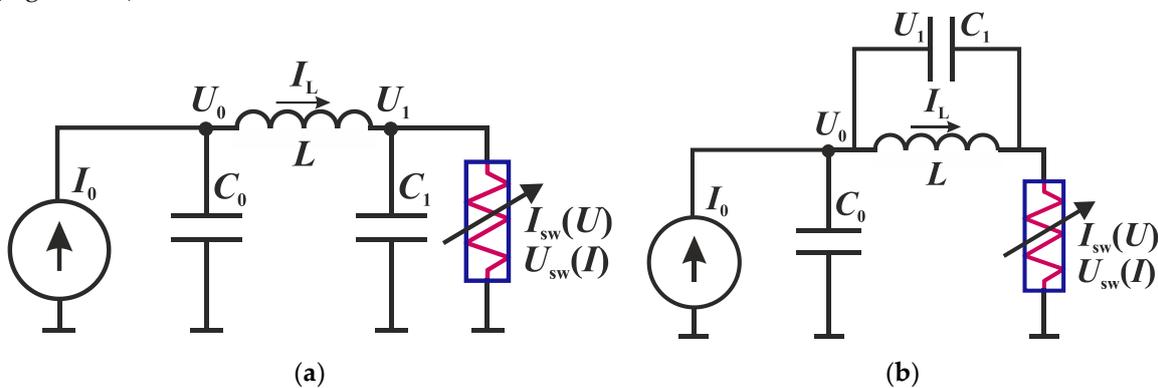

**Figure 11.** Neural-like circuits based on the S-type switch, generating bursts of oscillation, with a capacitance $C_1$ connected in parallel to the switching element (**a**) and in parallel to the inductance (**b**).

The circuits in Figure 11a,b are modeled by the following systems of differential equations, respectively:

$$C_0 \frac{dU_0}{dt} = I_0 - I_L(t),$$
$$C_1 \frac{dU_1}{dt} = I_L(t) - I_{sw}(U_1(t)), \quad (16)$$
$$L \frac{dI_L}{dt} = U_0(t) - U_1(t)$$

$$C_0 \frac{dU_0}{dt} = I_0 - I_{sw}(U_0(t) - U_1(t)),$$
$$C_1 \frac{dU_1}{dt} = I_{sw}(U_0(t) - U_1(t)) - I_L(t), \quad (17)$$
$$L \frac{dI_L}{dt} = U_1(t)$$

where $U_0(t)$ and $U_1(t)$ are the voltages on the capacitors $C_0$ and $C_1$, and $I_{sw}(U)$ is a piecewise linear approximation of the I–V characteristic from Equation (1) with the condition in Equation (2).

In contrast to the system of FitzHugh–Rinzel in Equation (9), Equations (16) and (17) use the direct $I_{sw}(U)$, and not the inverse I–V characteristic $U_{sw}(I)$, which is a two-valued, discontinuous function. Therefore, mathematical models of these circuits, most likely do not have an analytical solution, cannot be analyzed, and have only a numerical solution. However, these circuits are simpler than the FitzHugh–Rinzel schemes (Figure 7), contain fewer elements, and allow the generation of neural-like bursts.

Let us demonstrate the operation of the circuit shown in Figure 11b with the switching element at different temperatures. The temperature dependences of the threshold parameters of the I–V characteristic are listed in Table 1, the data correspond to the switching $VO_2$ element described in the previous study [46].

**Table 1.** Temperature dependences of the threshold parameters of the I–V characteristic.

| Temperature, °C | $U_{th}$, V | $U_h$, V | $R_{on}$, Ω | $R_{off}$, Ω | $U_{cf}$, V |
|---|---|---|---|---|---|
| 25 | 5.36 | 1.247 | 53 | 2550 | 0.955 |
| 40 | 4.052 | 0.93 | 55 | 2216 | 0.758 |
| 50 | 2.714 | 0.607 | 58 | 1726 | 0.502 |

Based on the data in Table 1, Figure 12 demonstrates the temperature-dependent I–V characteristic using Equations (1) and (2).

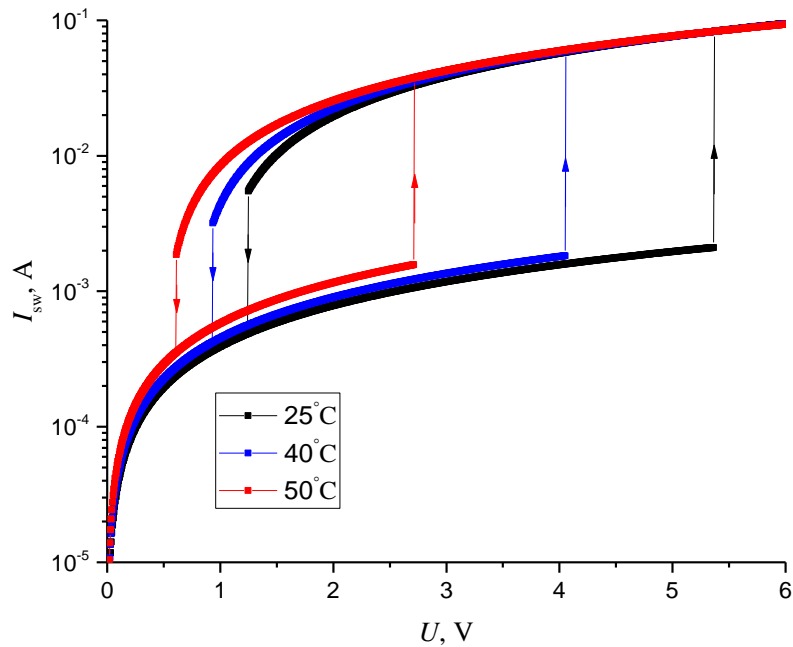

**Figure 12.** Temperature-dependent I–V characteristic, calculated by Equations (1) and (2) and the data in Table 1.

Modeled oscillograms of the current circuit (Figure 11b) at different temperatures of the $VO_2$ element are presented in Figure 13a. A pronounced burst mode can be observed at a temperature of 25 °C. As the temperature rises to 40 °C, a decrease from nine to three in the number of pulses in each pack and an increase in the frequency of repetition of the burst activity are visible. With a further increase in temperature to 50 °C, the bust oscillations become periodic single pulses with an increased frequency. Similar dynamics models the firing patterns of mammalian cold receptors (see Figure 13b) described in the previous study [47].

The external influence in the circuits can be implemented similarly to the FitzHugh–Nagumo circuit, by adding sources $U_b(t)$ and $I_b(t)$, and the interaction between the oscillators can be implemented through the thermal coupling, described in detail in the studies [12,48,49].

We have thus proposed the alternative neural-like schemes based on S-type switches. The $VO_2$ switch, due to its physical properties, can be used as a sensor model object to reproduce the pulse patterns of mammalian cold receptors.

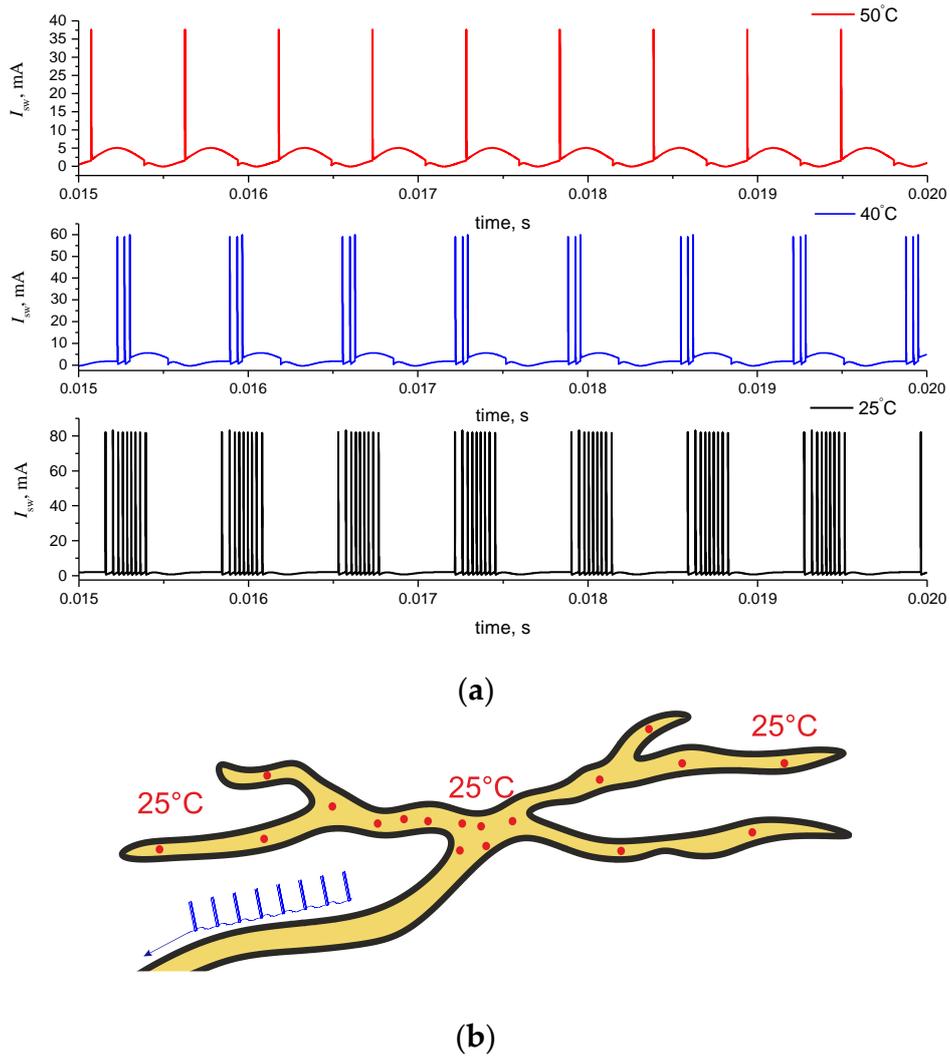

**Figure 13.** (**a**) Current oscillograms of the switching VO$_2$ element at different temperatures for the circuit in Figure 11b. Circuit parameters: $I_0$ = 2.5 mA, $C_0$ = 100 nF, $C_1$ =20 nF, and $L$ = 60 mH. (**b**) Schematic mapping of mammalian thermoreceptor at 25 °C, adopted from [50].

*3.4. The Auto-Relaxation Oscillator as an Integrate-And-Fire Neuron Based on a Switching S-Element*

When applying successive current pulses from various sources, the circuit of the relaxation generator described in Section 2.2 can simulate the operation of an integrate-and-fire neuron. Figure 14 captures an electronic circuit with two pulsed current sources $I_0$ and $I_1$. Figure 14b demonstrates an example of the operating point of the circuit in sub-threshold mode, when the current and voltage on the switch do not reach the threshold values ($U_{th}$, $I_t$). When the operating point is located on the high-resistance or low-resistance branch of the I–V characteristic, a stable state of the circuit is observed. When the operating point is transferred to the NDR region (for example, by changing the currents $I_0$ and $I_1$), the generation of spikes begins.

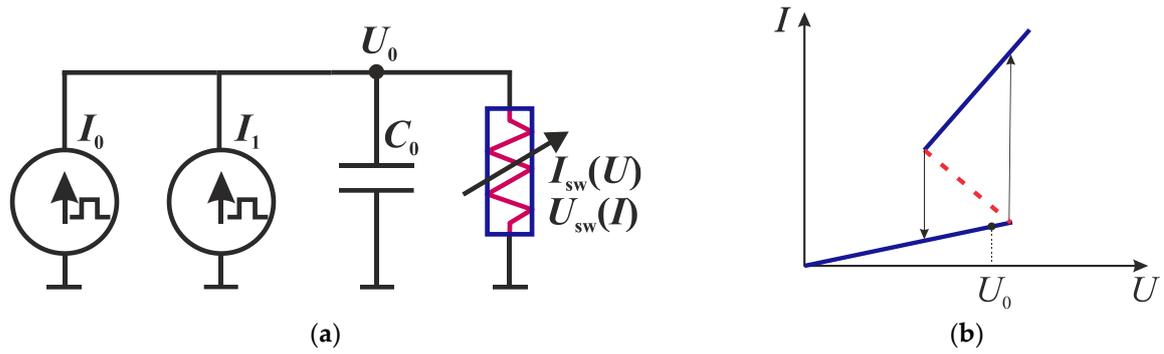

**Figure 14.** Circuit modeling the integrate-and-fire neuron (**a**) and an example of the operating point of the circuit in sub-threshold mode (**b**).

The circuit processing of two successive current pulses is reflected in the oscillogram (Figure 15), where $U_{sw}$ and $I_{sw}$ are the voltage and current on the switch, respectively. When the first current pulse $I_0$ is applied, the capacitor is charged to a voltage $U_0$, which does not reach the switch-on voltage of the switch $U_0 < U_{th}$. Then the discharge process of the capacitor starts through a switch with resistance $R_{off}$. When the voltage $U_{sw}$ drops to a certain value $U'_0$, the second current pulse $I_1$ is generated, the switch voltage rises to $U_{sw} = U_{th}$, and the switching process and the current pulse $I_{sw}$ are generated on the switch. Therefore, the first impulse sets the switch to the sub-threshold mode, when the operating point of the circuit is near the threshold (see Figure 14b), and the second impulse triggers the switch to turn on. Capacitor $C$ has the role of a current signal integrator, accumulating the charge, which increases the voltage on the capacitor, and it can ultimately lead to the generation of a signal at the output of the circuit.

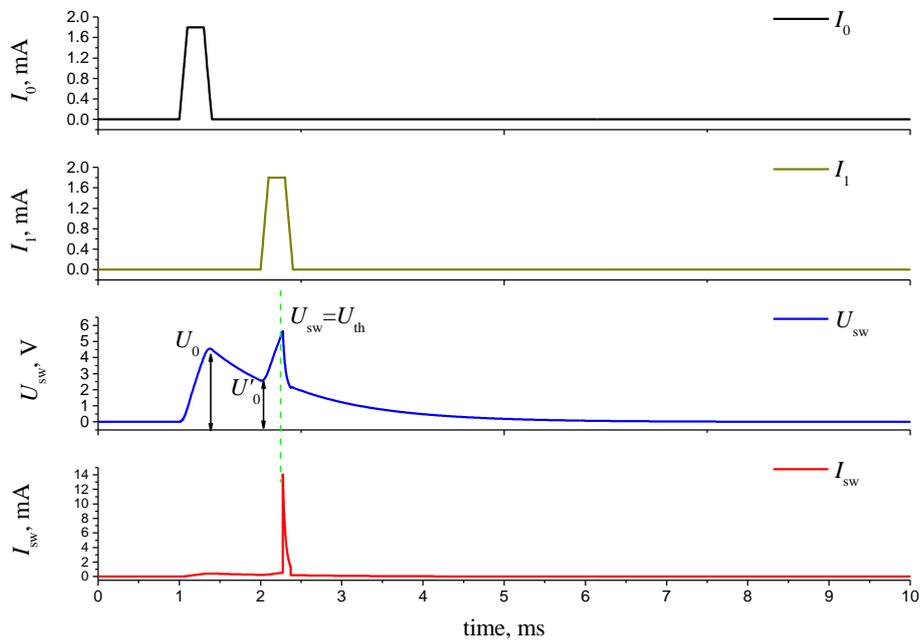

**Figure 15.** Oscillograms of the voltage $U_{sw}$ and the current $I_s$ response at the switch when exposed to two successive current pulses from sources $I_0$ and $I_1$. The circuit uses a VO$_2$ switch (Figure 2a).

## 4. Discussion

The circuit design of the FitzHugh–Nagumo model (Figure 6) can be compared with the well-known circuit on a tunnel diode [1,3], which has an N-type I–V characteristic. The main difference is that the tunnel diode is connected not in series, but in parallel to the inductance. Therefore, the voltage across the diode models the membrane potential $G(t)$, and the inductance current models the recovery potential $W(t)$ of the neuron.

A modification of the FitzHugh–Nagumo circuit, by adding an RC chain, derives the FitzHugh–Rinzel circuit model (Figure 9) based on an element with an S-shaped I–V characteristic. The transformation $(G, W, Q, \gamma) \rightarrow (-G, -W, -Q, -\gamma)$ leads the degenerate system in Equation (7) to its own form. However, in our FitzHugh–Rinzel model with a piecewise linear function shown in Equation (12) for positive values of $\gamma$, it is necessary to make another replacement $F_{pw}(G) \rightarrow -F_{pw}(-G)$ to get the same set of solutions. In the circuit, we change the direction of the supply current $I_0$ and use symmetrical with respect to the inversion $(I, U) \rightarrow (-I, -U)$ branch of the I–V characteristic of the switching S-element.

If—for obtaining relaxation oscillations—the parameters selection of the FitzHugh–Nagumo circuit does not have practical difficulties, the burst mode for the FitzHugh–Rinzel model (Figure 7) is implemented in a narrow range of parameters, and its search is not an easy task. The dynamics of the FitzHugh–Rinzel model with a nonlinear piecewise function has not yet been studied, and the values of bifurcations are not yet known. This could be a subject of scientific interest, both from a fundamental and practical point of view.

In the circuit design, the role of capacitors $C_0$ and $C_1$ in the FR circuits (Figure 9) can be explained as follows. The burst oscillation mode has two time parameters: Low-frequency, which corresponds to the oscillation period of the bursts, and high-frequency, which determines the period of pulses inside the bursts. The two capacitors $C_0$ and $C_1$ differ significantly in nominal value (by more than 15 times, see Figure 10 description). Large capacitance $C_0$, by gradually charging (from the source) and discharging (through the switch circuit), controls the low-frequency oscillation mode. The voltage $U_0(t)$ (see Figure 10b) changes in a narrow range and sets the operating point of the entire circuit. As a result, the switch either enters the oscillation mode or exits the oscillation mode, forming burst oscillations. The smaller capacitance $C_1$ controls the period of high-frequency pulsed oscillations inside the bursts. Similar considerations can be applied to alternative neural-like circuits (Figure 11).

The mandatory presence of the inductance in the proposed circuits ensures the correct functioning of the circuits only if the S-switch contains a stable NDR section (Figure 3). In practice, the chain elements always possess parasitic capacitances, and the switching would not happen instantaneously. Therefore, the FitzHugh–Nagumo and FitzHugh–Rinzel circuits, are probably capable of functioning with S-switches that have an unstable NDR (Figure 2), however, the possible effect of high-voltage induction on the inductive element may lead to the failure of the switch. In addition, the presence of a stable or unstable NDR is often determined by the manufacturing technology of the switching element. Therefore, the oxides of transition metals, similar to presented here $NbO_2$ and $VO_2$, can be probably used to create elements of both types.

Recent studies highlighted the interest of $VO_2$ switch applications in the SNN, ONN, and neural-like circuits [9,12,48,51–53]. The study [9] illustrates the implementation of $VO_2$ oscillators with various chaotic and burst modes of spike generation based on two switches. The circuits presented in this study that contain only one $VO_2$ switch have obvious advantages.

The alternative circuits, presented in Figure 11, have a simpler design and contain fewer elements than the FitzHugh–Rinzel circuits (Figure 7), however, they allow the generation of neural-like burst oscillations. The physical properties of the $VO_2$ switch enable the demonstration of the neuromorphic behavior of the circuit and the reproduction of the functioning of mammalian cold receptors.

The temperature dependence of the I–V characteristics (see Table 1) is attributed to the extremely strong influence of the temperature on the conductivity of the $VO_2$ film, as this material is used in the manufacturing of bolometric matrices [54] and temperature sensors [55]. A particularly strong dependence exists in the region of the metal to insulator phase transition (MIT), observed near the threshold temperature $T_t = 68$ °C, in the temperature hysteresis region. Above $T_t$, the $VO_2$ film operates in the high-conductivity (metal) phase, and, below $T_t$, the $VO_2$ film exists in the low-conductivity (insulator) phase. The jump in resistance between these phases can reach several orders of magnitude ($10-10^4$), depending on the structure of the film [55]. Many researchers, including us [56], have demonstrated that the effect of electrical switching occurs due to MIT, when the current passing through the VO$_2$ structure heats it with Joule heat to a temperature $T_t$. There is a strong dependence of the threshold switching parameters on the ambient temperature $T_0$ [46]. An increase in $T_0$ leads to a decrease in the threshold voltage $U_{th}$ and, at $T_0 \sim T_t$, the switching effect will be suppressed since the VO$_2$ channel will always be in a highly conductive state. This imposes a limitation on the use of a VO$_2$ sensor based on the effect of electrical switching. Most switching elements based on transition metal oxides have strong temperature dependences for $U_{th}$, and some structures, for example, based on NbO$_2$, demonstrate electrical switching up to temperatures of ~300 °C [24]. Therefore, the circuits presented in the current paper, based on elements with the S-shaped I–V characteristic, have the potential for practical application in a wide temperature range. In the future, assembling a neural network on such elements, using thermal [12,48] or electrical coupling between oscillators, it is possible to create systems with artificial intelligence that have temperature receptors.

The study results indicate that elements with the S-shaped I–V characteristic, having a wide variety of structures and materials, can form simple neural-like electronic circuits that only include one switch and passive elements. Switching properties and, primarily, the stability of switching parameters play a key role in the selection of S-elements suitable for the creation of such circuits. Currently, only silicon-type S-elements (trigger diodes), which are highly stable, are produced on an industrial scale. A number of researchers, including the authors of this article, have experimentally obtained the stable switches of planar and sandwich types [56]. Experimental samples of vanadium dioxide cannot yet compete with silicon elements, but the number of switching cycles can reach the values up to $10^{12}$. Pairing VO$_2$ switches into simple relaxation circuits (such as Figure 5a) in laboratory conditions is performed using probe stations, or by creating a cell with clamped or soldered contacts.

Spike and oscillatory neural networks open a promising direction in the development of neuroinformatics for the implementation of real-time behaving systems, and the manufacturing of very-large-scale integration circuits [17]. A significant advantage of pulsed signals is the availability of diverse methods for information encoding and synchronization. Methods of information encoding in ONN have been intensively developed, and the academic authors actively contribute to the development of new synchronization assessment techniques [12,16]. The simple burst oscillator circuits, presented in the current paper, can be of applied interest for the development of spike and oscillatory neural networks.

The FN and FR models are research subjects in the field of mathematical modeling of nonlinear dynamical systems. Despite the fact that the classical FN neural-like model (with cubic nonlinearity) has been well studied, its various modifications of the bi-stable mode, for example, with a piecewise linear function, may be of fundamental interest. Figure 6 can serve as a physical prototype of the model to check the dynamics and transitions between the modes of the oscillator operation. The same applies to the FR model, which is less studied and demonstrates an even more diverse dynamic mode.

Another prominent example of the practical application of neuromorphic circuits is the brain–machine interface. FN circuits are used to couple external electrical signals with living neurons [57] and to simulate the dynamics of neural activity [58]. In [57], an external electronic FN generator, which has a complex circuit with operational amplifiers, has been used to generate an optical signal that acted on the hippocampal living neurons, forming a neural interface for transmitting the effect. In this way, biosimilar electrical circuits can effectively couple electronics with living neurons, and, in the future, simple and reliable circuits, as proposed in this paper, can be utilized.

Circuit of the auto-relaxation generator that simulates the work of an integrate-and-fire neuron is interesting, as it operates with current pulses. The current pulses at the input are converted into a current pulse of the switch at the output. Such circuits can be connected using current repeaters or directly, through diode bridges and resistors. Creating an effective neural circuit on current pulses can be the subject of future research.

A detailed investigation and operation modes testing of the circuits presented in the current study can be performed using LTspice files (see Supplementary Materials).

## 5. Conclusions

In the current study, we proposed circuit solutions that implement the FitzHugh–Nagumo and FitzHugh–Rinzel models with the S-type switching element. Simplified circuits have been developed that allow modeling the integrate-and-fire neuron and the burst oscillation mode with emulation of mammalian cold receptor patterns.

The authors hope the results of this study would spark the interest of researchers in the experimental implementation of the proposed circuits, the mathematical analysis of dynamics of the systems with non-linear S-elements, and the development of oscillator and spike neural networks based on switching S-elements.

**Supplementary Materials:** The modelling code and instructions are available online at supplementary_materials.zip.

**Author Contributions:** Conceptualization P.B., A.V; methodology, P.B and A.V.; software validation, P.B and A.V.; writing—original draft preparation, P.B and A.V.; project administration, A.V.

**Funding:** This research was supported by the Russian Science Foundation (grant no. 16-19-00135).

**Acknowledgments:** The authors express their gratitude to Dr. Andrei Rikkiev for the valuable comments in the course of the article translation and revision.

**Conflicts of Interest:** The authors declare no conflict of interest.

**Appendix A**

The mathematical model of the circuit (Figure 9b) is as follows:

$$C_0 \frac{dU_0}{dt} = I_0 - \frac{U_0(t)}{R_0} - \frac{U_0(t) - U_1(t)}{R_1},$$

$$C_1 \frac{dU_1}{dt} = \frac{U_0(t) - U_1(t)}{R_1} - I_L(t), \quad \text{(A1)}$$

$$L \frac{dI_L}{dt} = U_1(t) - \left(U_b(t) + U_S(I_L(t) + I_{b0})\right)$$

Using the parameter $R_{mp}$ of S-type I–V characteristic in Equation (10) and dimensionless time $\tau \equiv t \cdot R_{mp}/L$, the first two equations of system in Equation (A1) are re-formulated in matrix form:

$$\begin{cases} \begin{pmatrix} \dfrac{dU_0}{d\tau} \\ \dfrac{dU_1}{d\tau} \end{pmatrix} = \dfrac{aR_{mp}}{R_1} \cdot \overbrace{\begin{pmatrix} -(1+b) & 1 \\ c & -c \end{pmatrix}}^{M} \times \begin{pmatrix} U_1 \\ U_2 \end{pmatrix} + a \begin{pmatrix} R_{mp}I_0 \\ -cR_{mp}I_L \end{pmatrix}, \\ \dfrac{d(R_{mp}I_L)}{d\tau} = U_1 - \left(U_b + U_{sw}(I_L + I_{b0})\right) \end{cases} \quad (18)$$

adding dimensionless circuit parameters:

$$a = \frac{L}{R_{mp}^2 C_0}, \quad b = \frac{R_1}{R_0}, \quad c = \frac{C_0}{C_1}. \quad \text{(A3)}$$

Linear transformation of the variables:

$$E_1(\tau) = \frac{c}{D} \cdot U_1(\tau) - \left(\frac{1}{2} - m\right) \cdot U_2(\tau),$$

$$E_2(\tau) = \frac{c}{D} \cdot U_1(\tau) + \left(\frac{1}{2} + m\right) \cdot U_2(\tau) \quad \text{(A4)}$$

$$m = \frac{b-c+1}{2D}, \quad D = \sqrt{(b-c+1)^2 + 4c},$$

where diagonalizes the matrix *M* of the first two equations of Equation (A2). As a result, Equation (A2) is reduced to the form:

$$\frac{dE_1}{dt'} = \frac{aR_{mp}}{R_1}\lambda_1 E_1 + ac\left(\frac{1}{2} - m\right) \cdot R_{mp}I_L + \frac{ac}{D} \cdot R_{mp}I_0,$$

$$\frac{dE_2}{dt'} = \frac{aR_{mp}}{R_1}\lambda_2 E_2 - ac\left(\frac{1}{2} + m\right) \cdot R_{mp}I_L + \frac{ac}{D} \cdot R_{mp}I_0, \quad (A5)$$

$$\frac{d(R_{mp}I_L)}{dt'} = E_2 - E_1 - (U_b + U_{sw}(I_L + I_{b0}))$$

where $\lambda_1 = -(D \cdot (m + \frac{1}{2}) + c)$ and $\lambda_2 = -(D \cdot (m - \frac{1}{2}) + c)$ are the characteristic numbers of the matrix *M*. Finally, a dimensionless change of variables is done in Equation (A5):

$$G(\tau) = \frac{I_L(\tau)}{I_{mp}}, \quad W(\tau) = \frac{1}{U_{mp}} \cdot \left(E_1(\tau) + \frac{cR_1I_0}{\lambda_1 D}\right),$$

$$Q(\tau) = \frac{E_2(\tau)}{U_{mp}}, \quad I_{ext}(\tau) = \frac{1}{U_{mp}} \cdot \left(\frac{cR_1I_0}{\lambda_1 D} - U_{in}(\tau)\right) \quad (A6)$$

a system, similar to the FitzHugh–Rinzel model, is then obtained:

$$\frac{dG}{dt'} = F_{pw}(G) - W + Q + I_{ext}$$

$$\frac{dW}{dt'} = \frac{aR_{mp}}{R_1}\lambda_1 W + ac\left(\frac{1}{2} - m\right) \cdot G, \quad (A7)$$

$$\frac{dQ}{dt'} = \frac{aR_{mp}}{R_1}\lambda_2 Q - ac\left(\frac{1}{2} + m\right) \cdot G + \frac{ac}{D} \cdot \frac{I_0}{I_{mp}}$$

where $F_{pw}(G)$ is the piecewise linear function

$$F_{pw}(G) = -\left(\frac{U_{b0}}{U_{mp}} + \frac{U_{sw}(G \cdot I_{mp} + I_{b0})}{U_{mp}}\right). \quad (A8)$$

The coefficients of the FitzHugh–Rinzel model in Equation (7) are determined from the parameters of the diagram in Figure 9b and from the system of Equation (A7) as follows:

$$\tau_0 = \frac{1}{ac\left(\frac{1}{2} - m\right)}, \quad \tau_1 = \frac{1}{ac\left(\frac{1}{2} + m\right)}, \quad \beta = \frac{-\lambda_1 R_{mp}}{cR_1\left(\frac{1}{2} - m\right)},$$

$$\chi = \frac{-\lambda_2 R_{mp}}{cR_1\left(\frac{1}{2} + m\right)}, \quad \gamma = \frac{I_0}{I_{mp}D\left(\frac{1}{2} + m\right)} \quad (19)$$

The characteristic numbers $\lambda_1$ and $\lambda_2$ of the matrix *M* are always negative, and the modulo of parameter *m* does not exceed ½. Therefore, all the parameters in Equation (A9) are always positive. For the constructed FitzHugh–Rinzel model with $\gamma > 0$, the piecewise linear function in Equation (A8) is the inverse of function in Equation (12) (see comments in the Discussion section).